\documentclass[12pt]{article}
\usepackage{epsfig}
\usepackage{epsfig}
\begin{document}
\def\tr{{\rm tr}\, }
\def\ints{\int_{S_2}}
\def\Tr{{\rm Tr}\, }
\def\hTr{\hat{\rm T}{\rm r}\, }
\def\be{\begin{eqnarray}}
\def\ee{\end{eqnarray}}
\def\ctt{\chi_{\tau\tau}}
\def\cta{\chi_{\tau a}}
\def\ctb{\chi_{\tau b}}
\def\cab{\chi_{ab}}
\def\cba{\chi_{ba}}
\def\bepsilon{\bar{\epsilon}}
\def\e{\epsilon}
\def\o{\over}
\def\ptt{\phi_{\tau\tau}}
\def\pta{\phi_{\tau a}}
\def\ptb{\phi_{\tau b}}
\def\>{\rangle}
\def\<{\langle}
\def\d{\hbox{d}}
\def\pab{\phi_{ab}}
\def\lb{\label}
\def\appendix{{\newpage\section*{Appendix}}\let\appendix\section%
        {\setcounter{section}{0}
        \gdef\thesection{\Alph{section}}}\section}
\renewcommand{\figurename}{Fig.}
\renewcommand\theequation{\thesection.\arabic{equation}}
\hfill{\tt }\\\mbox{} \vskip0.3truecm
\begin{center}
\vskip 2truecm {\Large\bf The volume of causal diamonds,
 asymptotically
\vskip 0.3truecm
 de Sitter
 space-times
and irreversibility}
 \vskip 1.5truecm {\large\bf Sergey
N.~Solodukhin\footnote{ {\tt solodukh@lmpt.univ-tours.fr}}
}\\
\vskip 0.6truecm \it{Laboratoire de Math\'ematiques et Physique
Th\'eorique CNRS-UMR 6083, \par Universit\'e de Tours, Parc de
Grandmont, 37200 Tours, France}
\end{center}
\vskip 1cm
\begin{abstract}
\noindent In this note we prove that the volume of a causal
diamond associated with an inertial observer in asymptotically de
Sitter 4-dimensional space-time is monotonically increasing
function of cosmological time. The asymptotic value of the volume
is that of in maximally symmetric de Sitter space-time. The
monotonic property of the volume is checked in two cases: in
vacuum and in the presence of a massless scalar field. In vacuum,
the volume flow (with respect to cosmological time) asymptotically
vanishes if and only if future space-like infinity is 3-manifold
of constant curvature. The volume flow thus represents
irreversibility of asymptotic evolution in spacetimes with
positive cosmological constant.
\end{abstract}
\vskip 1cm
\newpage

\vskip 1cm

\section{Introduction}
\setcounter{equation}0

In this paper we continue the study of geometry of causal diamonds
initiated in \cite{GS1} and \cite{GS2}. The focus of the present
study is the irreversible behavior of the volume of a causal
diamond in space-times asymptotic to  de Sitter space. The causal
diamonds play an important role in various recent classical and
quantum investigations of cosmologies with positive cosmological
constant (see \cite{B1} and \cite{B2} and references therein). In
fact, the diamond appears rather naturally as the region
accessible for experiments made by  a hypothetical observer moving
along a time-like geodesic. Imagine an observer that makes
experiments by sending the light rays and detecting the signals
that come back and has only a finite duration time $\tau$ for
his/her experiments. Then the region of space-time that can be
probed by this type of experiments is exactly the causal diamond
associated with the observer. Geometrically, as was demonstrated
in \cite{GS1} and \cite{GS2}, the volume of a causal diamond
encodes information on the curvature of the space-time. Moreover,
this information is inherently irreversible: the differences of
the space-time geometry inside the diamond from  de Sitter
geometry inevitably disappear as cosmological time progresses.

There are several indications in the literature that the
cosmological evolution with positive cosmological constant is
irreversible. The most straightforward  way is to
associate\footnote{I thank G. Gibbons for suggesting this point.}
this to the irreversible growth of the cosmological horizon
\cite{GH}. The entropy associated to the horizon is then
non-decreasing in agreement with the laws of thermodynamics.

In a different development one considers a dual holographic
description of de Sitter space-time in terms of a conformal field
theory (CFT)  \cite{St} defined on a space-like boundary of the
space-time. From the point of view of quantum CFT it is natural to
define a function (known as C-function) which changes
monotonically along the RG trajectories.  In the dual description
such a C-function is defined in terms of the space-time metric and
its derivatives. One then shows that, under suitable energy
conditions one has to impose on  matter fields, the C-function
changes monotonically with cosmological time \cite{St}.

In the present paper we suggest yet another manifestation of the
irreversible cosmological evolution. We show that the volume of a
causal diamond grows monotonically with cosmological time. The
maximal value it approaches at future space-like infinity is that
of volume in maximally symmetric de Sitter space-time.

\section{The result}
\setcounter{equation}0

We consider an observer that follows a timelike geodesic $\gamma$
in metric which is not exactly but only asymptotically de-Sitter,
in the limit that his/her  own proper time $t_q\rightarrow
\infty$. We shall study the volume $V(\tau, t_q)$ of the causal
diamond $\dot{I}^+(p) \cap \dot{I}^-(q)$ where $p$ and $q$ lie on
$\gamma$ in the limit when both $t_p, \ t_q \rightarrow \infty$
while $\tau=t_q-t_p$ is kept fixed. Thus both points $p$ and $q$
tend to future spacelike infinity ${\cal I}^+$ while the duration
of the diamond $\tau$ is kept fixed. The entire diamond is in the
asymptotic region and the volume of the diamond depends on the
asymptotic geometry. In the limit when $t_q\rightarrow \infty$ the
point $q$ on the geodesic $\gamma$ approaches the point $q^+$ of
intersection of geodesic $\gamma$ and the future space-like
infinity ${\cal I}^+$. The asymptotic metric in 4d  geodesic
coordinates takes the form \be
ds^2=-dt^2+e^{2t}g^{(0)}_{ij}dx^idx^j~~, \lb{m} \ee where $t$ is
the cosmological time and $g^{(0)}_{ij}(x)$ is an {\it arbitrary}
3d metric defined on   future infinity of the 4-dimensional
asymptotically de Sitter space-time.  We skip the subleading
terms, defined as a series in powers of $e^{-t}$, in (\ref{m}).
Throughout the paper we set the de Sitter radius $l=1$.

A special role is played by the  maximally symmetric de Sitter
space-time\footnote{Below we  call it  ``pure de Sitter
space-time''.}. This space-time is characterized by the fact that,
in global coordinates, the future infinity ${\cal I}^+$ is 3d
round sphere $S^3$, \be ds^2_{\rm
dS}=-dt^2+\cosh^2(t)\left(d\chi^2+\sin^2\chi(d\theta^2+\sin^2\theta
d\phi^2)\right)~~. \lb{global} \ee There are however two other
forms of the metric in the coordinates which cover only a part of
the space-time \be ds^2_{\rm
dS}=-dt^2+\exp(2t)\left(d\chi^2+\chi^2(d\theta^2+\sin^2\theta
d\phi^2)\right) \lb{met1} \ee and \be ds^2_{\rm
dS}=-dt^2+\sinh^2(t)\left(d\chi^2+\sinh^2
\chi(d\theta^2+\sin^2\theta d\phi^2)\right)~~. \lb{met2} \ee In
all three cases the  metric $g^{(0)}_{ij}$ on asymptotic boundary
is of constant curvature and it satisfies condition
$R^{(0)}_{ij}={1\over 3}g^{(0)}_{ij}R^{(0)}$. Notice that  not all
4d metrics which approach 3d manifold of constant curvature at
future infinity are globally identical to pure de Sitter
space-time.

The expansion of the volume $V(\tau, t_q)$ in powers of $e^{-t_q}$
is then a expansion in curvature of 3-dimensional Euclidean metric
$g^{(0)}_{ij}$ defined on ${\cal I}^+$ at point $q^+$ \be V(\tau,
t_q)=a_0(\tau)+a_2(\tau)R_{(0)}e^{-2t_q}+\left(c_4(\tau)\nabla^2R_{(0)}
+a_4(\tau)(R^{(0)}_{ij})^2+b_4(\tau)R_{(0)}^2\right)e^{-4t_q},
\lb{1} \ee where the curvature is taken at point $q^+$.
 There are no combinations of curvature which are odd in
 derivatives. This explains why the coefficients in front of odd powers of
 $e^{-t_q}$ in (\ref{1})
vanish.

One can easily get some constraints on the coefficients in
(\ref{1}). If the spacetime is the pure de Sitter spacetime, then
the volume of the causal diamond does not depend on where this
diamond is located. This is a direct consequence of the large
symmetry group in de Sitter space-time. It follows that the volume
$V_{\rm dS}(\tau, t_q)$ does not depend on $t_{q}$. The direct
calculation performed for metric in any form (\ref{global}),
(\ref{met1}) or (\ref{met2}) gives  \be V_{\rm dS}(\tau,
t_q)=v(\tau)\equiv {4\over 3}\pi \left(2\ln \cosh{\tau\over
2}-\tanh^2{\tau \over 2}\right)~~. \lb{v} \ee
 In this case all terms in the expansion (\ref{1})
except the first should vanish. For metric (\ref{global})  the
spacelike infinity ${\cal I}^+$ is   3-sphere with curvature
$R^{(0)}_{ij}=2g^{(0)}_{ij}$, $R_{(0)}=6$. Thus, we get that \be
a_0(\tau)=v(\tau)~,~~a_2(\tau)=0~,~~b_4(\tau)=-1/3 a_4(\tau)~~,
\lb{2} \ee where $v(\tau)$ is the volume of the diamond in pure de
Sitter spacetime. That coefficient $a_2(\tau)$ identically
vanishes was checked explicitly in  \cite{GS2}. Thus we have that
\be V(\tau, t_q)=v(\tau)+\left(c_4(\tau)\nabla^2R_{(0)}
+a_4(\tau)((R^{(0)}_{ij})^2-{1\over 3} R_{(0)}^2)
\right)e^{-4t_q}+.. ~~.\lb{3} \ee The functions $a_4(\tau)$ and
$c_4(\tau)$ can not be determined from general arguments and one
has to perform a direct calculation.

We use in ${\cal I}^+$ the
 Riemann coordinates  centered at point $q^+$ and directly compute
 the volume. The calculation shows that $c_4(\tau)=0$ identically
 and that $a_4(\tau)=-w(\tau)$ is entirely negative function of $\tau$,
\be V(\tau, t_q)=v(\tau)-w(\tau)\left(R^{(0)}_{ij}-{1\over
3}g^{(0)}_{ij}R^{(0)}\right)^2e^{-4t_q}+.. ~~.\lb{vt} \ee Thus,
the cosmological evolution defines a volume flow of a causal
diamond so that the volume is monotonically increasing function of
cosmological time. The asymptotic value of the volume is that of
in pure de Sitter spacetime. Moreover, the flow vanishes (to
leading order) if  future infinity ${\cal I}^+$ is 3-manifold of
constant curvature.

In the presence of 4d matter the monotonic behavior of the volume
of a causal diamond persists although the deviations from the pure
de Sitter result show up already in the second order in
$e^{-t_q}$. This is also obvious from the analysis similar to
(\ref{1}): one can use the asymptotic values of the matter fields
to construct new invariants that may appear in the expansion of
the volume together with the curvature invariants. For a massless
scalar field which takes value $\phi_0(x)$ on ${\cal I}^+$ the
asymptotic expansion of the volume takes the form \be V(\tau,
t_q)=v(\tau)-(2\pi G_N)c(\tau)(\nabla\phi_0)^2e^{-2t_q}+..
~~,\lb{vm} \ee where $G_N$ is 4d Newton's constant, $c(\tau)$ is
positive function of $\tau$ and
$(\nabla\phi_0)^2=g^{ij}_{(0)}\partial_i\phi_0\partial_j\phi_0$.

\bigskip

 Below we present a mathematical proof of  (\ref{vt}) and
(\ref{vm}).

\section{Asymptotic metric and the Riemann coordinates}
\setcounter{equation}0 We choose a time coordinate $\eta=e^{-t}$,
$\eta\geq 0$, $\eta=0$ at future infinity. Note that we are using
a convention in which $\eta$ is positive and decreases towards
future timelike infinity ${\cal I}^+$.

The asymptotic expansion of cosmological 4-metric with positive
cosmological constant was first considered by Starobinsky
\cite{Starobinsky}. It goes similarly to the expansion in
asymptotically anti-de Sitter case, see \cite{FeffermanGraham},
\cite{HS}, \cite{HSS} for more detail on the anti-de Sitter case.
The analytic continuation to the de Sitter case was considered in
\cite{Skenderis}, \cite{Anderson} and \cite{GS2}. The
4-dimensional metric takes the form \be &&ds^2={1\over
\eta^2}\left(-d\eta^2+g_{ij}(x,\eta)dx^idx^j\right)~~,\nonumber \\
&&g(x,\eta)=g^{(0)}(x)+g^{(2)}(x)\eta^2+g^{(3)}(x)\eta^3+g^{(4)}(x)\eta^4+..~~,
\lb{4} \ee where $\{x^i\}$ are  coordinates on ${\cal I}^+$. The
coefficients in the decomposition (\ref{4}) satisfy relations
\cite{Starobinsky} \be g^{(2)}_{ij}=R^{(0)}_{ij}-{1\over 4}R^{(0)}
g^{(0)}_{ij}~,~~\Tr g^{(3)}=0~,~~\nabla^j g^{(3)}_{ij}=0~,~~\Tr
g^{(4)}={1\over 4}\Tr g_{(2)}^2 ~~,\lb{5} \ee where the covariant
derivatives and trace are determined with respect to metric
$g^{(0)}_{ij}(x)$ defined on 3-surface ${\cal I}^+$.

Now on the surface ${\cal I}^+$ we choose the Riemann coordinates
$\{x^i\}$ such that $x^i=0$ correspond to point $q^+$. Locally,
around point $q^+$,  it is more convenient to  use the ``spherical
coordinates'' $(r,\theta^a)$ on ${\cal I}^+$,
 \be
x^i=rn^i(\theta)~,~~i=1,~2,~3 \ \ \ n^i(\theta)n^i(\theta)=1~~,
\lb{9} \ee where $\{\theta^a,a=1,2\}$ are the angle coordinates on
$S_2$.
 We then
develop  a  double expansion of metric both in  powers of $\eta$
and $r^2=x^ix^i$
 \be
&&g^{(0)}_{ij}=g^{(0,0)}_{ij}+g^{(0,2)}_{ij}r^2+g^{(0,3)}_{ij}r^3+g^{(0,4)}_{ij}r^4+..~~,\nonumber
\\
&&g^{(2)}=g^{(2,0)}+g^{(2,1)}r+g^{(2,2)}r^2+..~~,\nonumber \\
&&g^{(3)}=g^{(3,0)}+g^{(3,1)}r+..~~,\nonumber \\
&&g^{(4)}=g^{(4,0)}+.. ~~,\lb{6} \ee where we keep   terms up to
4th order in the total power of $\eta$ and $r$. In the Riemann
coordinates we have that \cite{Petrov} \be
&&g^{(0,0)}_{ij}=\delta_{ij}~~,\\
&&g^{(0,2)}_{ij}=-{1\over 3} R_{ikjn}n^kn^n~~,\nonumber \\
&&g^{(0,3)}_{ij}=-{1\over 6}R_{ikjn,l}n^k n^n n^l ~~,\nonumber \\
&&g^{(0,4)}_{ij}=(-{1\over 20}R_{ikjn,lm}+{2\over 45}R_{kin}^{\ \
\ \rho}R_{ljm\rho})n^kn^nn^ln^m~~,\nonumber \\
&&g^{(2,0)}_{ij}=(R_{ij}-{1\over 4} R\delta_{ij})~~,\nonumber \\
&&g^{(2,1)}_{ij}=\nabla_k(R_{ij}-{1\over 4}Rg_{ij})n^k~~,\nonumber \\
&&g^{(2,2)}_{ij}=\left({1\over 2}\nabla_k\nabla_n(R_{ij}-{1\over
4}Rg_{ij})-{1\over 6}R^\rho_{\ kin}(R_{\rho j}-{1\over 4}Rg_{\rho
j})-{1\over 6}R^\rho_{\ kjn}(R_{\rho i}-{1\over 4}Rg_{\rho
i})\right) n^k n^n~~.\nonumber \lb{7}\ee Expanding (\ref{5}) in
powers of $r$ we  get the relations \be \Tr g^{(3,0)}=\Tr
g^{(3,1)}=0~,~~\Tr g^{(4,0)}={1\over 4}\Tr g^2_{(2,0)} ~~.\lb{8}
\ee

\section{The future and past light-cones}
\setcounter{equation}0 We choose the  point $q$ to have
coordinates $(\eta=\epsilon, 0,0,0)$ and point $p$ to have
coordinates $(\eta=N+\epsilon,0,0,0)$, where
$N=\epsilon(e^\tau-1)$ and $\tau$ is the geodesic distance between
points $p$ and $q$, $\epsilon=e^{-t_q}$.

In  coordinates $(\eta, r,\theta^a)$ the equation which determines
the past  light-cone $\dot{I}^-(q)$, $r=r_+(\eta)$,
$\theta^a=const$ is \be &&{dr_+(\eta)\over d\eta}={1\over
\sqrt{g_{nn}}}~,~~g_{nn}=g_{ij}(x,\eta)n^in^j~,~~r_+(\eta=\epsilon)=0~~.
\lb{10}\ee We find that
 \be
 {1\over \sqrt{g_{nn}}}=1-{\eta^2\over
2}(g^{(2,0)}_{nn}+g^{(2,1)}_{nn} r+g^{(2,2)}_{nn}r^2)-{\eta^3\over
2}(g^{(3,0)}_{nn}+g^{(3,1)}_{nn}r)-{\eta^4\over
2}g^{(4,0)}_{nn}+{3\over 8}\eta^4(g^{(2,0)}_{nn})^2+.. \nonumber
\ee where we introduced notations
$g^{(k,p)}_{nn}=g^{(k,p)}_{ij}n^in^j$. The solution of  equation
(\ref{10}) including the terms up to 5th order takes the form \be
&&r_+(\eta)=\eta-\epsilon+{\epsilon^3\over 6}
g^{(2,0)}_{nn}+({g^{(3,0)}_{nn}\over 8}-{g^{(2,1)}_{nn}\over
24})\epsilon^4 +({g^{(2,2)}_{nn}\over 60}+{g^{(4,0)}_{nn}\over
10}-{g^{(3,1)}_{nn}\over 40}-{3\o
40}(g^{(2,0)}_{nn})^2)\epsilon^5\nonumber \\
&&-{\eta^3\o 6}(g^{(2,0)}_{nn}-\epsilon g^{(2,1)}_{nn}+\epsilon^2
g^{(2,2)}_{nn})-{\eta^4\o
8}(g^{(3,0)}_{nn}+g^{(2,1)}_{nn}-\epsilon g^{(3,1)}_{nn}-2\epsilon
g^{(2,2)}_{nn})\nonumber \\
 &&-{\eta^5\o
10}(g^{(2,2)}_{nn}+g^{(4,0)}_{nn}+g^{(3,1)}_{nn}-{3\o 4}
(g^{(2,0)}_{nn})^2)+.. ~~.\lb{11} \ee Similarly, the equation that
determines the light-cone $\dot{I}^+(p)$, $r=r_-(\eta)$,
$\theta^a=const$
 \be
{dr_-(\eta)\o d\eta}=-{1\o
\sqrt{g_{nn}}}~,~~r_-(\eta=N+\epsilon)=0~~. \lb{12} \ee
Introducing $\bar{\epsilon}=N+\epsilon$ we have that \be
&&r_-(\eta)=\bar{\epsilon}-\eta-{\bar{\epsilon}^3\over 6}
g^{(2,0)}_{nn}-({g^{(3,0)}_{nn}\over 8}+{g^{(2,1)}_{nn}\over
24})\bar{\epsilon}^4 -({g^{(2,2)}_{nn}\over
60}+{g^{(4,0)}_{nn}\over 10}+{g^{(3,1)}_{nn}\over 40}-{3\o
40}(g^{(2,0)}_{nn})^2)\bar{\epsilon}^5\nonumber \\
&&+{\eta^3\o 6}(g^{(2,0)}_{nn}+\bar{\epsilon}
g^{(2,1)}_{nn}+\bar{\epsilon}^2 g^{(2,2)}_{nn})+{\eta^4\o
8}(g^{(3,0)}_{nn}-g^{(2,1)}_{nn}+\bar{\epsilon}
g^{(3,1)}_{nn}-2\bar{\epsilon}
g^{(2,2)}_{nn})\nonumber \\
 &&+{\eta^5\o
10}(g^{(2,2)}_{nn}+g^{(4,0)}_{nn}-g^{(3,1)}_{nn}-{3\o 4}
(g^{(2,0)}_{nn})^2)+.. ~~.\lb{13} \ee Two light-cones,
$\dot{I}^+(p)$ and $\dot{I}^-(q)$, intersect at \be
\eta=\eta_c\equiv{N\o 2}+\epsilon-{1\o 8}g^{(2,0)}_{nn}N^2({N\o
2}+\epsilon )+O(\epsilon^4) ~~.\lb{14} \ee

\section{The volume}
\setcounter{equation}0 The volume of the causal diamond
$\dot{I}^+(p)\cap \dot{I}^-(q)$  is \be V(\epsilon,
\tau)=\int_{S_2}\left(\int_\epsilon^{\eta_c}{d\eta\o \eta^4}
\int_0^{r_+(\eta)}dr\ r^2 \ \sqrt{\det
g}+\int_{\eta_c}^{N+\epsilon}{d\eta\o \eta^4}\int_0^{r_-(\eta)}dr\
r^2 \ \sqrt{\det g}\right)~~, \lb{15} \ee where $\int_{S_2}$ is
the integral over spherical angles $\{\theta^a,\ a=1,2\}$. With
the usual choice of angles $\int_{S_2}=\int_0^\pi
d\theta\sin\theta\int_0^{2\pi}d\phi$.

Then we get for the past light-cone of point $q$
 \be \int_0^{r_+(\eta)}dr\ r^2 \sqrt{\det
g}={1\o 3}(\eta-\epsilon)^3+S^+_5+S^+_6+S^+_7 ~~,\lb{17-1} \ee
where the exact form of the coefficients $S^+_5, S^+_6$ is given
in Appendix A. The coefficient $S^+_7$ takes the form
 \be
S^+_7=\sum_{n=0}^4h^+_n\eta^n(\epsilon-\eta)^{7-n} ~~, \lb{20-1}
\ee where $h^+_n, \ n=0, 1, 2, 3, 4$ are presented in Appendix A.

For the future light-cone of point $p$ we have that
 \be
\int_0^{r_-(\eta)}dr\ r^2 \sqrt{\det g}={1\o
3}(\bar{\epsilon}-\eta)^3+S^-_5+S^-_6+S^-_7 ~~.\lb{26-1} \ee The
exact form of $S^-_5$ and $S^-_6$ can be found in Appendix A.
Using exact expressions of Appendix A we  notice a property \be
S^-_5(\bepsilon,\eta)=-S^+_5(\e=\bepsilon,\eta) ~~.\lb{29-1}\ee
For the term $S^{-}_7$ we find \be
 S^-_7=\sum_{n=0}^4h^-_n\eta^n(\bepsilon-\eta)^{7-n}~~, \lb{30-1} \ee
where $h^-_n, \ n=0, 1, 2, 3, 4 $ are  given by relation
(\ref{hh}). Using the $S_2$ integrals calculated  in Appendix B,
we get for the integrated quantities
 \be
&&\int_{S_2}h_4^+=0~~,\\
&&\int_{S_2}h_3^+=-{B\o 3}+{5C\o 48} ~~,\nonumber \\
&&\int_{S_2}h_2^+={4C\o 45}-{B\o 3}~~, \nonumber \\
&&\int_{S_2}h_1^+={5C\o 144}-{B\o 6}+{A\o 120}~~,\nonumber \\
&&\int_{S_2}h_0^+=-{113 B\o 3150}+{227 C\o 37800}+{A\o 280}~~,
\nonumber \lb{31}\ee where we introduced $A\equiv\pi \nabla^2 R$,
$B\equiv\pi R_{ij}^2$, $C\equiv\pi R^2$.

We notice that since $\int_{S_2}g^{(3,1)}_{nn}=0$ one has a
relation \be
\int_{S_2}h^{-}_n=-\int_{S_2}h^{+}_n~,~~n=0,1,2,3,4~~. \lb{hm} \ee
Now we are in a position to calculate the volume of the causal
diamond. We focus on the term proportional to $\epsilon^4$ since
all other terms, proportional to $\epsilon^2$ and $\e^3$ vanish.
First, we neglect the modification (\ref{14}) and assume that two
light-cones intersect  at $\eta_c={N\o 2}+\e$. We then get \be
V^{(4)}_1=\int_\e^{{N\o 2}+\e}{d\eta\o
\eta^4}S^+_7(\e,\eta)+\int_{{N\o 2}+\e}^{N+\e}{d\eta\o
\eta^4}S_7^-(\bar{\e},\eta)=\sum_{n=0}^4h^{+}_nI_n(K)
\epsilon^4~~, \lb{32}\ee where \be I_n(K)=\int_{1}^K dx
(1-x)^{7-n}\left({1\o x^{4-n}}+{1\o (x-2K)^{4-n}}\right)\lb{33}\ee
and we introduced $K={N\o 2\e}+1={{e^\tau+1} \o 2}$. There are
useful relations between functions $I_n(K)$ \be &&I_2(K)=-{1\o
3}I_1(K)-{(1-K)^6\o
3K^2}~~,\nonumber \\
&&I_3(K)=-{1\o 5}I_2(K)~~,\nonumber \\
&&I_1(K)=-{3\o 7}I_0(K)~~. \lb{34} \ee Now we can take into
account the modification  (\ref{14}) of the intersection point of
the two light cones, \be \eta_c={N\o 2}+\e+\Delta
\eta_c~,~~\Delta\eta_c=-{1\o 2}g^{(2,0)}_{nn}K(K-1)^2\e^3~~.
\lb{35}\ee We get the contribution to the volume due to this
modification \be &&V^{(4)}_2=\int_{S_2}\int_{{N\o 2}+\e}^{{N\o
2}+\e+\eta_c} {d\eta\o \eta^4}\left({(\eta-\e)^3\o
3}-{(\bar{\e}-\eta)^3\o 3}+{1\o
2}g^{(2,0)}_{nn}\eta[(\eta-\e)^4+(\bar{\e}-\eta)^4]+..\right)\nonumber\\
&&=-{1\o 4}\int_{S_2}(g^{(2,0)}_{nn})^2{(K-1)^6\o
K^2}\e^4+O(\e^5)~~.
 \lb{36}\ee

The total volume then \be
V^{(4)}=V^{(4)}_1+V^{(4)}_2=\e^4\left({2(K-1)^6\o 135 K^2}+{4\o
4725} I_0(K)\right) (C-3B)~~. \lb{37} \ee

Notice that terms proportional to $A=\nabla^2 R$ cancel each
other. It is crucial for  establishing the monotonic behavior of
the volume  because the term $\nabla^2 R$ is not sign-definite. On
the other hand, one has that
$$(C-3B)=\pi(-3(R^{(0)}_{ij})^2+R_{(0)}^2)=-3\pi(R^{(0)}_{ij}-{1\o 3}R^{(0)}g^{(0)}_{ij})^2~~.$$
The integral $I_0(K)$ is calculated explicitly. Since
$K={{e^\tau+1}\o 2}$ one has that \be &&\left({2(K-1)^6\o 135
K^2}+{4\o 4725} I_0(K)\right)\equiv{1\o 3\pi}w(\tau)\nonumber \\
&&={2\o 135}(1-\tanh{\tau\o 2})^{-4}({1\o 15}\tanh^2{\tau \o
2}(\tanh^4{\tau\o 2}+12\tanh^2{\tau\o 2}+{6\o \cosh^4{\tau\o
2}}+{53\o \cosh^2{\tau\o 2}}+251)\nonumber \\
&&-2(1+\tanh{\tau\o 2})^4\ln (1+\tanh{\tau\o 2}) -2(1-\tanh{\tau\o
2})^4\ln (1-\tanh{\tau\o 2})) ~~.\lb{38} \ee The function
$w(\tau)$ is positive and rapidly growing with $\tau$. Finally, we
get for the volume of the causal diamond \be V(\tau,
t_q)=v(\tau)-w(\tau)(R^{(0)}_{ij}-{1\o
3}R^{(0)}g^{(0)}_{ij})^2e^{-4t_q}~~, \lb{39} \ee where $v(\tau)$
is the volume in maximally symmetric de Sitter spacetime. Thus, we
get that \be {dV(\tau, t_q) \o dt_q}=4w(\tau)(R^{(0)}_{ij}-{1\o
3}R^{(0)}g^{(0)}_{ij})^2e^{-4t_q}>0 ~~.\lb{40} \ee So that the
volume of the causal diamond  monotonically grows. The derivative
(\ref{40}) vanishes if curvature of ${\cal I}^+$ satisfies
relation $(R^{(0)}_{ij}={1\o 3}R^{(0)}g^{(0)}_{ij}) $. By Bianchi
identities this implies that $R^{(0)}=const$ so that ${\cal I}^+$
is 3-manifold of constant curvature in this case.

\section{Coupling to massless scalar field}
\setcounter{equation}0 In the presence of matter fields the
behavior (\ref{39}) changes in that the correction term to the
pure de Sitter result appears now at a different power of
$e^{-t}$. In order to illustrate this point we consider a massless
scalar field described by the field equation
 \be
{1\o \sqrt{G}}\partial_\mu
(\sqrt{G}G^{\mu\nu}\partial_\nu\phi)=0~~, \lb{Phi} \ee where
$G_{\mu\nu}$ is 4-metric.
 The
Einstein equations take the form \be R_{\mu\nu}=3G_{\mu\nu}+8\pi
G_N
\partial_\mu\phi\partial_\nu \phi~~. \lb{s1} \ee
For the 4-metric in the form (\ref{4}), introducing a coordinate
$\rho=\eta^2$, the Einstein equations reduce to a system of
equations \cite{HSS} \be \rho \,[2 g^{\prime\prime} - 2 g^\prime
g^{-1} g^\prime + \Tr\, (g^{-1} g^\prime)\, g^\prime]_{ij} &+&
R_{ij}(g) - (d - 2)\, g^\prime_{ij} - \Tr \,(g^{-1} g^\prime)\,
g_{ij} =
\nonumber \\
 &=&8\pi G_N\partial_i\phi\partial_j\phi~~, \nonumber \\
 \nabla_i\, \Tr \,(g^{-1}
g^\prime) - \nabla^j g_{ij}^\prime &=& 8\pi
G_N\partial_\rho\phi\partial_i\phi~~, \nonumber\\
 \Tr \,(g^{-1}
g^{\prime\prime}) - \frac{1}{2} \Tr \,(g^{-1} g^\prime g^{-1}
g^\prime) &=&  16\pi G_N\partial_\rho\phi\partial_\rho\phi ~~,
\label{s2} \ee where differentiation with respect to $\rho$ is
denoted with a prime, $\nabla_i$ is the covariant derivative
constructed from the metric $g$, and $R_{ij} (g)$ is the Ricci
tensor of $g$.

The asymptotic expansion for the scalar field and the 4-metric
reads \be \phi(\rho,x)=\phi^{(0)}(x)+\phi^{(2)}(x)\rho+..~~, \nonumber \\
g_{ij}(\rho,x)=g^{(0)}_{ij}(x)+g^{(2)}_{ij}(x)\rho+..~~. \lb{exp}
\ee Inserting this into the first equation in (\ref{s2}) we find
that \be g^{(2)}_{ij}=(R^{(0)}_{ij}-{1\o
4}g^{(0)}_{ij}R^{(0)})-8\pi
G_N(\partial_i\phi^{(0)}\partial_j\phi^{(0)}-{1\o
4}g^{ij}_{(0)}\partial_i\phi^{(0)}\partial_j\phi^{(0)}) \lb{g2}\ee
and, hence, for the trace \be \Tr g^{(2)}={1\o 4}R^{(0)}-2\pi
G_Ng^{ij}_{(0)}\partial_i\phi^{(0)}\partial_j\phi^{(0)}~~.
\lb{tg2}\ee

Before computing the volume of the causal diamond we notice that
all our expressions obtained in the previous section  and in
Appendix A and that operate with coefficients in the expansion
(\ref{6})  and do not refer to the precise form of the
coefficients are also valid in the case when the bulk gravity
couples to matter. We  also note that regardless to the precise
form of the coefficient $g^{(2)}_{ij}$ we have that \be
\int_{S_2}g^{(2,0)}_{nn}={1\o 3}\int_{S_2}\Tr
g^{(2,0)}~~.\lb{**}\ee

The volume can be computed in the same way as in the previous
section. The only difference is that now the term of order
$\epsilon^2$ does not identically vanish. We get that \be
V(\tau,t_q)=v(\tau)+\int_{S_2}\left(\int_\epsilon^{{N\o
2}+\epsilon}{d\eta\o \eta^4}S^+_5(\epsilon,\eta)+\int_{{N\o
2}+\epsilon}^{N+\epsilon}{d\eta\o
\eta^4}S^-_5(\bepsilon,\eta)\right)+..~~. \lb{00} \ee Inserting
here the exact expressions for $S^+_5$ and $S^-_5$ we arrive at
the expression \be V(\tau,\epsilon)=v(\tau)+\left({1\o
2}\int_{S_2}g^{(2,0)}_{nn}J_1(\tau)+({1\o 10}\int_{S_2}\Tr
g^{(0,2)}-{1\o
6}\int_{S_2}g^{(2,0)}_{nn})J_2(\tau)\right)\epsilon^2+..
~~,\lb{44} \ee where we introduced functions \be
J_1(\tau)=\int_1^Kdx (x-1)^4\left({1\o x^3}+{1\o (x-2K)^3}\right)
\lb{J1} \ee and \be
 J_2(\tau)=\int_1^Kdx
(x-1)^5\left({1\o x^4}+{1\o (x-2K)^4}\right)~~, \lb{J2} \ee where
$K={e^\tau+1\o 2}$. We note a relation between two functions \be
J_2(\tau)={5\o 3}J_1(\tau)~~. \ee Taking into account that
$g^{(0,2)}_{ij}=-{1\o 3}R_{ikjn}n^kn^n$ we find  \be \int_{S_2}\Tr
g^{(0,2)}=-{4\pi\o 9}R_{(0)}~~,\nonumber \\
\int_{S_2}g^{(2,0)}_{nn}={4\pi\o 3}({1\o 4}R_{(0)}-2\pi G_N
(\nabla_{(0)}\phi_{(0)})^2)~~, \lb{88} \ee where all quantities
are calculated at point $q^+$ on ${\cal I}^+$.

Thus,  we obtain for the volume
 \be
 V(\tau,t_q)=v(\tau)-{8\pi\o 27}J_1(\tau)(2\pi
 G_N)(\nabla_{(0)}\phi_{(0)})^2e^{-2t_q}+..~~.
 \lb{VV}\ee
 The function $J_{1}(\tau)$ is positive and rapidly growing with
 $\tau$,
 \be
 J_1(\tau)={\tau^6\o 320}+{\tau^7\o 320}+{19\o
 17920}\tau^8+O(\tau^9)~~,~~~\tau \ll 1 \\
 J_1(\tau)=({17\o 4}-6\ln 2)e^{2\tau}-{3\o
 2}e^\tau+O(1)~~,~~~\tau \gg 1
 \lb{J11}
 \ee
Clearly, the volume (\ref{VV}) is monotonically growing. It
reaches asymptotically the maximal value that is the volume of the
diamond in pure de Sitter space-time. The volume flow (\ref{VV})
vanishes everywhere in ${\cal I}^+$ if and only if the scalar
field takes a constant asymptotic value $\phi_{(0)}$ on ${\cal
I}^+$.

\section{Two conjectures}
\setcounter{equation}0

It  is reasonable to ask whether  the volume of a causal diamond
is globally monotonic or, rather, it is monotonic only
asymptotically. It is known \cite{Friedrich} (see also
\cite{Anderson}) that there exists a small perturbation of  pure
de Sitter space-time that does not change the global structure of
the space-time. It makes a small deformation of past and future
infinities ${\cal I}^-$ and ${\cal I}^+$.
 Near the past
infinity our analysis is valid by replacing $t$ by $-t$. So that
the volume is monotonically decreasing near ${\cal I}^-$. Thus in
a space-time which is globally asymptotically de Sitter, to the
past and future, ${\cal I}^-$ and ${\cal I}^+$, the volume of a
causal diamond is {\it not} globally monotonic. On the other hand,
a positive energy distribution generically causes a formation of a
initial singularity. The space-time near the past infinity is then
no more asymptotically de Sitter. We expect that in  space-times
of this type the volume of the diamond is globally monotonic as a
manifestation of the irreversibility of the cosmological evolution
with a positive cosmological constant. A detail analysis, however,
is yet to be done.

Irrespective of whether or not the volume is globally monotonic it
is reasonable to expect that  the volume of a diamond in pure de
Sitter space-time is the absolute maximum  among all possible
asymptotically de Sitter space-times. Thus, we formulate two
conjectures:

{\it 1. The volume of a causal diamond in pure de Sitter
space-time is the absolute maximum in the class of vacuum
 asymptotically de Sitter metrics} \be
V(\tau, t_q)\leq V_{\rm dS}(\tau) ~~.\lb{con1} \ee
\medskip

{\it 2. The bound (\ref{con1}) is saturated for all diamonds of
same duration $\tau$  if and only if the space-time is pure de
Sitter space-time.}

\medskip

A more work is needed to check these statements.

\bigskip

It should be noted that the assumption of an asymptotically de
Sitter space-time at $t\rightarrow \infty$ is crucial for all
results obtained in the paper.  The only known generic alternative
behaviour is recollapse with the subsequent formation of a
singularity. We do not expect the volume of a diamond to be
monotonic in this scenario although a detail analysis is needed in
order to specify the time evolution of the volume.

\bigskip

\bigskip

\bigskip

\noindent {\large \bf Acknowledgments} \\
\vskip 2mm \noindent I thank G. Gibbons for  the fruitful
collaboration on \cite{GS1} and \cite{GS2} and for useful
comments.

\bigskip

\noindent

\appendix{Volume coefficients}
\setcounter{equation}0 Using the decomposition \be&&\sqrt{\det
g}=1+{r^2\o 2}\Tr g^{(0,2)}+{r^3\o 2}\Tr g^{(0,3)}+{r^4\o 2}\Tr
g^{(0,4)}+{1\o 2}\eta^2(\Tr
g^{(2,0)}+r\Tr g^{(2,1)}+r^2\Tr g^{(2,2)})\nonumber \\
&&+{1\o 8}(r^2\Tr g^{(0,2)}+\eta^2\Tr g^{(2,0)})^2-{r^4\o
4}\Tr(g^{(0,2)}g^{(0,2)})-{r^2\eta^2\o
2}\Tr(g^{(0,2)}g^{(2,0)})-{\eta^4\o 8}\Tr(g^{(2,0)}g^{(2,0)})+..
\nonumber \\
\lb{16} \ee we get \be \int_0^{r_+(\eta)}dr\ r^2 \sqrt{\det
g}={1\o 3}(\eta-\epsilon)^3+S^+_5+S^+_6+S^+_7 ~~,\lb{17} \ee where
\be && S^+_5= {1\o 6}(\Tr
g^{(2,0)}-3g^{(2,0)}_{nn})\eta^2(\eta-\epsilon)^3+{1\o
2}g^{(2,0)}_{nn}\eta(\eta-\epsilon)^4\lb{18}\\
&&+({1\o 10}\Tr g^{(0,2)}-{1\o 6}g^{(2,0)}_{nn})(\eta-\epsilon)^5
\nonumber \ee

\be &&S^+_6=-{g^{(3,0)}_{nn}\o
2}\epsilon^3(\eta-\epsilon)^3+(-{g^{(2,1)}_{nn}\o 4}-{3\o
4}g^{(3,0)}_{nn}+{\Tr g^{(2,1)}\o
8})\epsilon^2(\eta-\epsilon)^4 \\
&&+(-{g^{(2,1)}_{nn}\o 3}-{g^{(3,0)}_{nn}\o 2}+{\Tr g^{(2,1)}\o
4})\epsilon(\eta-\epsilon)^5+({\Tr g^{(2,1)}\o 8}+{\Tr g^{(0,3)}\o
12}-{g^{(3,0)}_{nn}\o 8}-{g^{(2,1)}_{nn}\o 8})(\eta-\e )^6
\nonumber \lb{19} \ee

For the coefficient $S^+_7$ we get a representation
 \be
S^+_7=\sum_{n=0}^4h^+_n\eta^n(\epsilon-\eta)^{7-n} \lb{20} \ee

\be h^+_4={1\o 2}g^{(4,0)}_{nn}-{5\o 8}(g^{(2,0)}_{nn})^2-{1\o
24}(\Tr g^{(2,0)})^2+{1\o 4} g^{(2,0)}_{nn}\Tr g^{(2,0)}+{1\o
24}\Tr(g^{(2,0)}g^{(2,0)}) \lb{21} \ee

\be h^+_3=-{5\o 4}(g^{(2,0)}_{nn})^2+{1\o 4}g^{(2,0)}_{nn}\Tr
g^{(2,0)}+g^{(4,0)}_{nn} -{1\o 4}g^{(3,1)}_{nn}\lb{22} \ee

\be && h^+_2={1\o 4}g^{(2,0)}_{nn}\Tr g^{(0,2)}-{7\o
6}(g^{(2,0)}_{nn})^2-{1\o 10}\Tr g^{(2,2)}+{1\o 10}\Tr
(g^{(0,2)}g^{(2,0)})-{1\o 20}\Tr g^{(2,0)}\Tr g^{(0,2)} \nonumber
\\
&& +{1\o 6}g^{(2,2)}_{nn}+g^{(4,0)}_{nn}+{1\o 12}g^{(2,0)}_{nn}\Tr
g^{(2,0)}-{1\o 4}g^{(3,1)}_{nn} \lb{23} \ee

\be h^+_1={1\o 4}g^{(2,0)}_{nn}\Tr g^{(0,2)}-{13\o
24}(g^{(2,0)}_{nn})^2+{1\o 12}g^{(2,2)}_{nn}+{1\o
2}g^{(4,0)}_{nn}-{1\o 8}g^{(3,1)}_{nn} \lb{24} \ee

\be &&h^+_0=-{37\o 360}(g^{(2,0)}_{nn})^2-{1\o 56}(\Tr
g^{(0,2)})^2+{1\o 28}\Tr (g^{(0,2)}g^{(0,2)})\nonumber \\
&&-{1\o 14}\Tr g^{(0,4)}+{1\o 60}g^{(2,2)}_{nn}+{g^{(4,0)}_{nn}\o
10}+{1\o 12}g^{(2,0)}_{nn}\Tr g^{(0,2)} -{1\o
40}g^{(3,1)}_{nn}\lb{25} \ee

For the future cone of point $p$ we have that
 \be
\int_0^{r_-(\eta)}dr\ r^2 \sqrt{\det g}={1\o
3}(\bar{\epsilon}-\eta)^3+S^-_5+S^-_6+S^-_7 \lb{26} \ee

\be && S^-_5= -{1\o 6}(\Tr
g^{(2,0)}-3g^{(2,0)}_{nn})\eta^2(\eta-\bepsilon)^3-{1\o
2}g^{(2,0)}_{nn}\eta(\eta-\bepsilon)^4\\
&&-({1\o 10}\Tr g^{(0,2)}-{1\o
6}g^{(2,0)}_{nn})(\eta-\bepsilon)^5\nonumber \lb{27} \ee

\be &&S^+_6={g^{(3,0)}_{nn}\o
2}\bepsilon^3(\eta-\bepsilon)^3+(-{g^{(2,1)}_{nn}\o 4}+{3\o
4}g^{(3,0)}_{nn}+{\Tr g^{(2,1)}\o
8})\bepsilon^2(\eta-\bepsilon)^4 \\
&&+(-{g^{(2,1)}_{nn}\o 3}+{g^{(3,0)}_{nn}\o 2}+{\Tr g^{(2,1)}\o
4})\bepsilon(\eta-\bepsilon)^5+({\Tr g^{(2,1)}\o 8}+{\Tr
g^{(0,3)}\o 12}+{g^{(3,0)}_{nn}\o 8}-{g^{(2,1)}_{nn}\o 8})(\eta-\e
)^6 \nonumber \lb{28} \ee

Clearly, we have a property \be
S^-_5(\bepsilon,\eta)=-S^+_5(\e=\bepsilon,\eta) ~~.\lb{29}\ee For
the term $S^{-}_7$ we find \be
 S^-_7=\sum_{n=0}^4h^-_n\eta^n(\bepsilon-\eta)^{7-n}~~,  \lb{30} \ee
where we have a relation \be &&h^-_4=-h^+_4~~, \lb{hh} \\
&&h^-_3=-h^+_3-{1\over 2}g^{(3,1)}_{nn}~~, \nonumber \\
&&h^-_2=-h^+_2-{1\over 2}g^{(3,1)}_{nn}~~,\nonumber \\
&&h^-_1=-h^+_1-{1\over 4}g^{(3,1)}_{nn}~~,\nonumber \\
&&h^-_0=-h^+_0-{1\over 20}g^{(3,1)}_{nn}~~.\nonumber  \ee Since
$\int_{S_2}g^{(3,1)}_{nn}=0$ we have that \be
\int_{S_2}h_n^-=-\int_{S_2}h_n^+~~,~~n=0,\ 1,\ 2,\ 3,\ 4 \lb{hint}
\ee

\appendix{Spherical integrals}
\setcounter{equation}0 Calculating integrals over spherical angle
coordinates we use that \be &&\int_{S_2}n^in^j={4\o 3}\pi
\delta^{ij}~~,\nonumber
\\
&&\int_{S_2}n^i n^j n^k n^l={4\o 15}\pi
(\delta^{ij}\delta^{kl}+\delta^{ik}\delta^{jl}+\delta^{il}\delta^{jk})~~.
\ee

We  use the fact that in three dimensions \be
R_{ijkn}=g_{ik}P_{jn}+g_{jn}P_{ik}-g_{jk}P_{in}-g_{in}P_{jk}~,~~P_{ij}=R_{ij}-{1\o
4}g_{ij}R~~.
 \ee

We then get, introducing $A\equiv\pi \nabla^2 R$, $B\equiv\pi
R_{ij}^2$, $C\equiv\pi R^2$,

\be &&\int_{S^2}g^{(3,1)}_{nn}=0 ~~,\\
&&\ints g^{(2,2)}_{nn}={A\o 10} ~~,\nonumber \\
&&\ints g^{(4,0)}_{nn}={B\o 3}-{5C\o 48} ~~,\nonumber \\
&&\ints (g^{(2,0)}_{nn})^2={8B\o 15}-{3C\o 20}~~,\nonumber \\
&&\ints g^{(2,0)}_{nn}\Tr g^{(2,0)}={C\o 12}~~,\nonumber \\
&&\ints g^{(2,0)}_{nn}\Tr g^{(0,2)}={C\o 45}-{8B\o 45}~~,\nonumber \\
&&\ints(\Tr g^{(2,0)})^2={C\o 4}~~,\nonumber \\
&&\ints\Tr g^{(2,0)}\Tr g^{(0,2)}=-{C\o 9}~~,\nonumber \\
&&\ints \Tr g^{(0,2)}\Tr g^{(0,2)}={4C\o 135}+{8B\o 135}~~,\nonumber \\
&&\ints \Tr g^{(2,2)}={A\o 6}-{4B\o 9}+{C\o 9}~~,\nonumber \\
&&\ints \Tr g^{(0,4)}=-{2A\o 75}+{56B\o 675} -{4C\o 225}~~,\nonumber \\
&&\ints \Tr(g^{(0,2)}g^{(2,0)})=-{4B\o 9}+{C\o 9} ~~,\nonumber \\
&&\ints \Tr(g^{(2,0)}g^{(2,0)})=4B-{5C\o 4}~~,\nonumber \\
&&\ints \Tr(g^{(0,2)}g^{(0,2)})={28B\o 135}-{2C\o 45}~~.
\nonumber\ee


\newpage



\begin{thebibliography} \\
\bibitem{GS1}
  G.~W.~Gibbons and S.~N.~Solodukhin,
  ``The geometry of small causal diamonds,''
  Phys.\ Lett.\  B {\bf 649}, 317 (2007); arXiv:hep-th/0703098.

\bibitem{GS2}
  G.~W.~Gibbons and S.~N.~Solodukhin,
  ``The Geometry of Large Causal Diamonds and the No Hair Property of
  Asymptotically de-Sitter Spacetimes,''
  Phys.\ Lett.\  B {\bf 652}, 103 (2007);
  arXiv:0706.0603 [hep-th].



\bibitem{B1}
  R.~Bousso,
  ``Positive vacuum energy and the N-bound,''
  JHEP {\bf 0011}, 038 (2000);
  arXiv:hep-th/0010252.


\bibitem{B2}
  R.~Bousso, R.~Harnik, G.~D.~Kribs and G.~Perez,
  ``Predicting the Cosmological Constant from the Causal Entropic Principle,''
  Phys.\ Rev.\  D {\bf 76}, 043513 (2007);
  arXiv:hep-th/0702115.


\bibitem{GH}
  G.~W.~Gibbons and S.~W.~Hawking,
  ``Cosmological Event Horizons, Thermodynamics, And Particle Creation,''
  Phys.\ Rev.\  D {\bf 15}, 2738 (1977).




\bibitem{St}
  A.~Strominger,
  ``Inflation and the dS/CFT correspondence,''
  JHEP {\bf 0111}, 049 (2001);
  arXiv:hep-th/0110087.



  \bibitem{Starobinsky}
A.~ A.~ Starobinsky, ``Isotropization of arbitrary cosmological
expansion given an effective cosmological constant'',  {\it JETP
Letters } {\bf  37} ( 1983)  66-69.

\bibitem{FeffermanGraham} C.~ Fefferman and C.~ R.~  Graham,
1985 Conformal invariants. In: {\it Elie Cartan et les
mathematiques d'aujourd'hui} . Asterisque (hors serie),(1985)
95-116.


\bibitem{HS}
  M.~Henningson and K.~Skenderis,
  ``The holographic Weyl anomaly,''
  JHEP {\bf 9807}, 023 (1998);
  arXiv:hep-th/9806087.


\bibitem{HSS}
  S.~de Haro, S.~N.~Solodukhin and K.~Skenderis,
Holographic reconstruction of spacetime and renormalization in the
AdS/CFT correspondence,
 {\it  Commun. Math. Phys.}  {\bf 217} (2001) 595;
  arXiv:hep-th/0002230.

\bibitem{Skenderis}
  K.~Skenderis,
  ``Lecture notes on holographic renormalization,''
  Class.\ Quant.\ Grav.\  {\bf 19}, 5849 (2002);
  arXiv:hep-th/0209067.

\bibitem{Petrov} A.~Z.~ Petrov,   "Einstein spaces" , Pergamon
(1969).

\bibitem{Friedrich} H.~Friedrich,
``On the existence of n-geodesically complete or future complete
solutions of Einstein's field equations with smooth asymptotic
structure,'' Comm.\ Math.\ Phys.\ {\bf 107}, 587-609 (1986).

\bibitem{Anderson}
  M.~T.~Anderson,
  ``Existence and stability of even dimensional asymptotically de Sitter
  spaces,''
  Annales Henri Poincare {\bf 6}, 801 (2005);
  arXiv:gr-qc/0408072; M.~T.~Anderson,
  ``On the structure of asymptotically de Sitter and anti-de Sitter spaces,''
  arXiv:hep-th/0407087.






\end{thebibliography}
\end{document}